\documentclass[10pt]{iopart}
\expandafter\let\csname equation*\endcsname\relax

\expandafter\let\csname endequation*\endcsname\relax
\usepackage{amsmath, amsthm, amssymb, amsfonts}
\usepackage{graphicx}
\usepackage{subcaption}
\usepackage{tabularx}
\usepackage{slashed}
\usepackage{xfrac}
\usepackage{multirow}
\usepackage{iopams}  
\usepackage{hyperref}
\hypersetup{colorlinks,
citecolor=blue, urlcolor=blue, linkcolor=blue}
\usepackage{upgreek}
\usepackage[greek,english]{babel}
\usepackage[utf8]{inputenc}
\usepackage{orcidlink}

\begin{document} 
\title{Backreactions from loading the stable photon sphere in Weyl conformal gravity}

\author{Reinosuke Kusano$^1$\orcidlink{0000-0001-7569-7197}, Keith Horne$^1$\orcidlink{0000-0003-1728-0304}, Friedrich Koenig$^1$\orcidlink{0000-0002-2250-2058}, Miguel Yulo Asuncion$^{1, 2, 3}$\orcidlink{0009-0003-5075-3107}} 
\address{$^1$SUPA Physics and Astronomy, North Haugh, University of St\,Andrews, KY16 9SS, Scotland, United Kingdom} 
\address{$^2$Nottingham Centre of Gravity, Nottingham NG7 2RD, United Kingdom} 
\address{$^3$School of Mathematical Sciences, University of Nottingham,
University Park, Nottingham NG7 2RD, United Kingdom} 
\ead{rk77@st-andrews.ac.uk} 
\vspace{10pt} 
\begin{indented} 
\item[]December 2025 
\end{indented}

\begin{abstract}
We investigate the accumulation of null matter at the stable photon sphere in the Mannheim-Kazanas metric, the analogue to the Schwarzschild solution  in Weyl's conformal theory of gravity. In our toy problem in which we consider an infinitely-thin shell, we find that a jump in radial pressure ${T^r}_r$ is induced across the shell unless the shell has a radius of either the unstable or stable photon sphere radii. We then find that upon loading the stable photon sphere, its area remains invariant. Furthermore, at a critical threshold loading limit for this zero-width null matter shell, we are able to produce a metric containing an extremal horizon with an AdS$_2\times$S$^2$ geometry completely independent of the cosmological curvature. This hitherto unencountered and therefore unexpected result is a phenomenon unseen in standard nonconformal second-order metrics with nonzero cosmological constants. 

\vspace{1pc} 
\end{abstract}

\submitto{\CQG}
%
%
%

\section{\label{sec:intro}Introduction} 

Albert Einstein's general theory of relativity (GR)~\cite{einstein1914} has persisted as the most important theory of gravitation in the last century. Starting with the perihelion precession of Mercury~\cite{einstein1915}, GR has had great success in predicting the gravitational redshift of photons~\cite{popper1954}, the gravitational lensing of light often used in the detection of astronomical objects~\cite{dyson1920}, and gravitational waves from binary black hole mergers~\cite{centrella2010}. 

Adopting the ``mostly-positive'' metric signature $(-,+,+,+)$, as well as natural units such that $c=\hbar=1$ here and henceforth, the gravitational action in GR is defined by the Einstein-Hilbert action $I_{\rm EH}$~\cite{hilbert1915}: 
\begin{equation}\label{eq:EH_action}
    I_{\rm EH} = -\dfrac{1}{16\pi G} \int \text{d}^4x (-g)^{\frac{1}{2}} (\mathcal{R}-2\Lambda),
\end{equation} 
where $G$ is the Newtonian gravitational constant, $g_{\mu\nu}$ is the metric tensor of which $g$ is the determinant, $\mathcal{R}$ is the Ricci scalar, and $\Lambda$ is the cosmological constant. Through the extremisation of the action with respect to the metric tensor, we obtain Einstein's famous field equations~\cite{einstein1914}: 
\begin{equation}\label{eq:EFE} 
    G_{\mu \nu} =- 8\pi GT_{\mu \nu} - \Lambda g_{\mu\nu},
\end{equation} 
with the Einstein tensor $G_{\mu \nu}$ and $T_{\mu \nu}$ the symmetric energy-momentum tensor. 

Despite the myriad successful predictions made by GR, significant inadequacies persist. Flat galaxy rotation curves~\cite{rubin1970, ciotti2022} and mass discrepancies in galaxy clusters~\cite{zwicky1940, rood1970} require the invocation of dark matter, an ad-hoc-seeming invisible quantity that is currently only indirectly detectable~\cite{alam2021, planck2015}. GR also suffers from the dark energy problem, where the vacuum energy measured from cosmological observables disagrees with quantum field theory predictions by more than 120 orders of magnitude~\cite{weinberg1989}. Additionally, while the strong, electromagnetic, and weak forces are conformally invariant, GR does not follow the same invariance, making quantizing GR as a quantum gravity (QG) theory undesirable~\cite{mannheim2006}. 

These tensions in GR are the main motivators of research into alternative theories of gravity; Weyl's conformal theory of gravity (CG) is one such theory. 

Following the relativity principle proffered by Weyl~\cite{weyl1918} and Bach~\cite{bach1921} at the start of the 20th century, CG was established by Mannheim and Kazanas in 1989 as a promising alternative to GR~\cite{mannheim1989}. While Einstein's field equations for GR are invariant under coordinate and Lorentz transformations, CG's field equations possess an additional invariance under Weyl transformations, which are local isotropic stretches of the metric preserving angles and shapes~\cite{mannheim1989}: 
\begin{equation}
    g_{\mu\nu}(x) \rightarrow \Omega^2(x)g_{\mu\nu}(x) ,
\end{equation} 
where $\Omega(x)$ is a conformal factor defining the stretch of the metric.

CG successfully fits flat galaxy rotation curves~\cite{deliduman2020, dutta2018, nesbet2015, obrien2012} and addresses the dark energy problem~\cite{mannheim2011} without reliance on the dark sector, as well as being able to reproduce GR results on solar system scales~\cite{sultana2012}. On the quantum gravity front, CG is a promising QG candidate due to it sharing the conformal invariance of the other forces of the Standard Model~\cite{mannheim2006}.

CG's gravitational action, the Weyl action $I_{\rm W}$, takes a different form to that of the Einstein-Hilbert action~\eqref{eq:EH_action}: 
\begin{equation}\label{eq:Weyl_action}
    I_{\rm W} = -\alpha_\mathrm{g} \int \text{d}^4 x (-g)^\frac{1}{2} C_{\lambda \mu \nu \kappa}C^{\lambda \mu \nu \kappa}.
\end{equation} 
Here, $\alpha_\mathrm{g}$ is a negative and dimensionless constant associated with the gravitational field~\cite{mannheim2007_2}, while the conformally invariant Weyl tensor $C_{\lambda \mu \nu \kappa}$ is the traceless part of the Riemann tensor $R_{\lambda \mu \nu \kappa}$~\cite{mannheim1989}: 
\begin{align}\label{eq:Weyl_tensor}
C_{\lambda \mu \nu \kappa} = R_{\lambda \mu \nu \kappa} -& \dfrac{1}{2}(g_{\lambda\nu} R_{\mu\kappa} - g_{\lambda\kappa}R_{\mu\nu} \notag - g_{\mu\nu}R_{\lambda\kappa}+g_{\mu\kappa}R_{\lambda\nu}) \notag \\&+ \dfrac{1}{6}{\mathcal{R}}(g_{\lambda\nu}g_{\mu\kappa} - g_{\lambda\kappa}g_{\mu\nu}),
\end{align}
with $R_{\mu\nu}$ denoting the Ricci tensor. In the action~\eqref{eq:Weyl_action}, $C_{\lambda \mu \nu \kappa}$ is squared as $C_{\lambda \mu \nu \kappa}C^{\lambda \mu \nu \kappa}$ to generate a scalar quantity with dimensions of length$^{-4}$ required in the integrand of the Lagrangian density~\eqref{eq:Weyl_action}. 

While GR's Einstein field equations in~\eqref{eq:EFE} are second-order, CG's conformally invariant field equations contain up to fourth derivatives of the metric, making CG a fourth-order theory of gravity:
\begin{equation}\label{eq:BE}
    4 \alpha_\mathrm{g} W_{\mu \nu} = T_{\mu \nu} ,
\end{equation}
where the Bach tensor $W_{\mu\nu}$, the CG analogue of the Einstein tensor $G_{\mu\nu}$, is given by: 
\begin{equation}\label{eq:bach_tensor}
    W_{\mu\nu}=W^{(2)}_{\mu\nu}-\dfrac{1}{3}W^{(1)}_{\mu\nu},
\end{equation}
with the functions $W_{\mu\nu}^{(1)}$ and $W_{\mu\nu}^{(2)}$ defined as~\cite{mannheim1989}:
\begin{equation}
    \begin{aligned}
        \begin{cases}
            W^{(1)}_{\mu\nu}&=2g_{\mu\nu}{\mathcal{R}^{;\lambda}}_{;\lambda}-2\mathcal{R}_{;\mu;\nu}-2\mathcal{R}R_{\mu\nu}+\dfrac{1}{2}\mathcal{R}^2,\\
            W^{(2)}_{\mu\nu}&=\dfrac{1}{2}g_{\mu\nu}{\mathcal{R}^{;\lambda}}_{;\lambda}+{{R_{\mu\nu}}^{;\lambda}}_{;\lambda}-{{R_{\mu}}^\lambda}_{;\mu;\lambda}-2R_{\mu\lambda}{R_{\nu}}^{\lambda}+\dfrac{1}{2}g_{\mu\nu}R_{\rho\lambda}R^{\rho\lambda}.
        \end{cases}
    \end{aligned}
\end{equation}
$W_{\mu\nu}$ is traceless, due to the conformal invariance of this gravitational theory. Subsequently,~\eqref{eq:BE} requires the energy-momentum tensor $T_{\mu\nu}$ to be traceless as well.

While higher-order derivative theories usually face scrutiny due to a purported violation of unitarity and the presence of ghosts~\cite{aspects}, specifically fourth-order derivative theories, such as CG, have been shown to preserve unitarity and ghostlessness~\cite{mannheim2007_1, bender2008}. This proof is attained by assuming that the quantum mechanical Hamiltonian does not need to be Dirac-Hermitian, but is instead $\mathcal{PT}$-symmetric. This allows for the construction of a fourth-order derivative model of the Pais-Uhlenbeck oscillator that is both unitary and ghost-free.

Additionally, a reassessment of the Ostrogradsky instability for higher-order derivative theorems found that the associated ghost is in fact avoidable. This becomes clear when one considers that the original Ostrogradsky theorem only necessarily holds true for Lagrangians of one dynamical variable, while modern field theories have multiple fields~\cite{ostrogradsky}.

\subsection{The Mannheim-Kazanas metric}\label{subsec:MK_metric}

CG is a metric-based theory of gravity, and it possesses black hole solutions much like GR~\cite{mannheim1989, mannheim1991_1}. CG's vacuum metric for a nonrotating and uncharged source is the Mannheim-Kazanas (MK) metric, the CG equivalent to the GR Schwarzschild (GRS) metric~\cite{mannheim1989}. We note that while some works call this the Mannheim-Kazanas metric after the authors of~\cite{mannheim1989}, this metric was already discovered half a decade earlier by Riegert~\cite{riegert1984}. Nevertheless, we refer to the CG Schwarzschild metric as the Mannheim-Kazanas (MK) metric throughout, to be consistent with~\cite{horne2016}. 

Both GRS and MK metrics may produce a nonrotating, uncharged black hole in spherical symmetry, described by a metric tensor of the form: 
\begin{align}\label{eq:sph_symm_BH} 
    g_{\mu\nu}=\text{diag}\left(g_{tt}, g_{rr}, g_{\theta\theta}, g_{\phi\phi}\right)=\text{diag}\left(-B(r), \dfrac{1}{B(r)}, r^2, r^2\sin^2\theta\right),
\end{align} 
such that the line element $\mathrm{d}s$ is given by:
\begin{equation}\label{eq:line_elem}
    \mathrm{d}s^2 = -B(r)\mathrm{d}t^2 + \dfrac{\mathrm{d}r^2}{B(r)} + r^2 \ {\mathrm{d}\Omega_2}^2,
\end{equation} 
where $B(r)$ is the blackening factor, and ${\mathrm{d}\Omega_2}^2$ is the 2-sphere metric ${\mathrm{d}\Omega_2}^2=\mathrm{d}\theta^2+\sin^2\theta \ \mathrm{d}\phi^2$, with $t$, $r$, $\theta$, and $\phi$ corresponding to time, radius, polar angle, and azimuthal angle in four-dimensional spacetime. The blackening factor of the GRS metric is given by~\cite{schwarzschild1916}:
\begin{equation}\label{eq:GRS}
    B(r)=1-\dfrac{2\beta}{r},
\end{equation}
where $\beta$ is the gravitational radius, or half the Schwarzschild radius associated with the mass~\cite{mannheim2011_2, kusano2025_CGRN, candlish2018}. For CG's equivalent to the GRS metric in~\eqref{eq:GRS}, $B(r)$ is defined through the fourth-order Poisson equation~\cite{mannheim2006}: 
\begin{equation}\label{eq:Poisson_4} 
    \nabla^4 B(r) = \dfrac{3 ({T^0}_0 - {T^r}_r)}{4\alpha_\mathrm{g} B(r)}=\dfrac{-3 (\rho + p_r)}{4\alpha_\mathrm{g} B(r)} \equiv f(r),
\end{equation} 
where $f(r)$ is a source function encoding energy density $\rho$ and radial pressure $p_r$ from the energy-momentum tensor. This source function is analogous to the mass density in the second-order Poisson equation. In addition to~\eqref{eq:Poisson_4}, we impose a third-order constraint from the $rr$ component of~\eqref{eq:BE}: 
\begin{equation}\label{eq:BE_rr}
    4 \alpha_\mathrm{g} {W^r}_r = {T^r}_r = p_r(r),
\end{equation}
with ${W^r}_r$ explicitly expressed as~\cite{horne2016, brihaye2009}: 
\begin{align}\label{eq:W^r_r}
    {W^r}_r&=\dfrac{1-B^2}{3r^4}+\dfrac{2BB'}{3r^3}-\dfrac{BB'' + (B')^2}{3r^2}+\dfrac{B'B''-BB'''}{3r}+\dfrac{2B'B'''-(B'')^2}{12},
\end{align} 
where primes denote derivatives in $r$.~\eqref{eq:Poisson_4} and~\eqref{eq:BE_rr} are therefore the two independent equations of motion to be solved for the blackening factor in the metric tensor. The metric Ansatz provided in~\eqref{eq:line_elem} restricts only the diagonal components of the Bach equations to be nonzero, and spherical symmetry imposes that $W^{\phi}_{\phi}=W^{\theta}_{\theta}$; we can always express two of the Bach equations~\eqref{eq:BE} in terms of the other two. For this reason, to generate a blackening factor fulfilling the Bach equations~\eqref{eq:BE}, we first solve~\eqref{eq:Poisson_4} and impose the constraint from the third-order equation presented in~\eqref{eq:BE_rr}.

In the sourceless case when $f(r)=0$, we have a homogeneous fourth-order Poisson equation $\nabla^4B(r)=0$, which is solved to give: 
\begin{equation}\label{eq:MK_metric}
    B(r) = w-\dfrac{2M}{r}+\gamma r -\kappa r^2.
\end{equation} 
If in a vacuum with $T_{\mu\nu}=0$, one may take the auxiliary parameters $w$ and $M$ to be: 
\begin{align}\label{eq:MK_params_aux}
    w &\equiv 1 - 3\beta\gamma, \hspace{6ex}
    M \equiv \beta\left(1 - \dfrac{3}{2}\beta\gamma\right).
\end{align}
$\beta$, $\gamma$, and $\kappa$ are constants of integration~\cite{mannheim1989}, which we refer to collectively as the MK parameters. In a more general sourceless case, assuming that $T_{\mu\nu}\neq0$ but still ${T^0}_0={T^r}_r$ such that the source function $f(r)=0$, we may simplify~\eqref{eq:BE_rr} to obtain the following relation between $w$, $M$, and $\gamma$, to reduce one degree of freedom in our choice of parameters~\cite{horne2016, brihaye2009}: 
\begin{equation}\label{eq:3_order}
    w^2=1-6M\gamma-\dfrac{3r^4}{4\alpha_\mathrm{g}}{T^r}_r,
\end{equation}
where ${T^{r}}_r$ is the radial pressure $p_r(r)$, which again in the sourceless case is equal to ${T^0}_0$. The MK parameters in~\eqref{eq:3_order} remain constant if $p_r\propto 1/r^4$, while the vacuum MK parameters of~\eqref{eq:MK_params_aux} fulfill the third-order constraint of~\eqref{eq:BE_rr} with $4\alpha_\mathrm{g}{W^r}_r={T^r}_r=0$. 

The GRS metric is recoverable by setting $\gamma,\kappa=0$. Setting $\gamma=0$ and $\kappa>0$ ($\kappa<0$) gives the GR Schwarzschild (Anti-)de Sitter metric (GRS(A)dS), meaning that we can associate $\kappa$ with the cosmological constant $\Lambda$ from~\eqref{eq:EFE} as $\kappa=\Lambda/3$~\cite{horne2006, gibbons2008}. The linear $\gamma r$ has been used to fit the flat rotation curves of galaxies~\cite{horne2016, hobson2021, mannheim2012}. 

Geometric features of nonrotating spacetime metrics include horizons where $B(r)=0$, as well as photon spheres, which are circular null geodesics around the central singularity. These are located at radii where $\mathrm{d}V_\mathrm{eff}/\mathrm{d}r=0$, and $V_\mathrm{eff}=B(r)/r^2$ (see section~\ref{subsec:ps_conditions}), allowing for stable or unstable photon orbits~\cite{qiao2022}. The GRS metric has one horizon at $r=2\beta$ and one unstable photon sphere at $r=3\beta$. On the other hand, the MK metric has a maximum of three horizons and two photon spheres, with one unstable photon sphere at $r_\text{ust}=3\beta$ analogous to the one in GRS, and one stable photon sphere at $r_\text{st}=3\beta-2/\gamma$~\cite{turner2020}. In particular, the stability of the latter suggests the likelihood for the buildup of null matter at these radii, which we may refer to as ``accumulation points''~\cite{turner2020}. As proven in~\cite{cunha2017}, photon spheres always come in pairs; naturally, as $\gamma=0$ in GRS, the stable photon sphere in this metric can be thought to be located at $r\rightarrow+\infty$. 

In the present work, we solve a toy problem to analyse photon accumulation in CG, by building up an infinitely-thin shell at the radius of the stable photon sphere $r=r_\mathrm{st}=3\beta-2/\gamma$. Previous works have attempted the investigation of self-gravitating photons at the photon sphere in the GRS metric~\cite{diFilippo2025}; investigating this previously-unexplored phenomenon with the MK metric proves to be more physically feasible, as the photons are not required to teeter exactly on the unstable extremum of the effective potential. As has also been seen in similar analyses of GR metrics~\cite{diFilippo2025}, such an investigation would prove to elucidate the stability of the MK metric when subject to backreactions from accumulating null matter.

We structure the present work as follows. In section~\ref{sec:MK_metric_pert}, we establish important terminology in discussing the causal structure of the MK metric, and introduce a screening method for finding physically feasible radii at which we may construct the photon sphere. We then describe the main findings of the present work in sections~\ref{sec:geometries_pert} to~\ref{sec:threshold&above}. In section~\ref{sec:geometries_pert}, using the fourth-order Poisson equation from~\eqref{eq:Poisson_4}, we analyse the requirements for thin-shell sources in the MK metric, and in section~\ref{subsec:horizons-pert} we visualise the differences between vacuum and thin-shell spacetimes. In section~\ref{sec:threshold&above}, we discuss the maximum load permitted by such a structure. We summarise our findings in section~\ref{sec:conclusion}, and suggest avenues for further developments.

\section{Spacetimes in the Mannheim-Kazanas metric}\label{sec:MK_metric_pert} 

This section is subdivided into three parts. Section~\ref{subsec:regions} introduces the two main spacetime regions of static nonrotating metrics, and section~\ref{subsec:horizons} introduces the different types of horizons which demarcate these regions. With regards to these definitions, in section~\ref{subsec:ps_conditions}, we establish selection procedures for appropriate parameter values for which a shell of null particles may exist in the MK metric. 

We conduct all further analysis with dimensionless parameters $r/\beta$, $\beta\gamma$, and $\beta^2\kappa$. We also restrict ourselves to ``positive mass'' solutions by setting $\beta>0$, as $\beta<0$ corresponds to central sources composed of exotic matter~\cite{turner2020}. 

\subsection{Spacetime regions}\label{subsec:regions} 

In the present work, we refer to two types of spacetime regions. In regions where $B(r)>0$, $\mathrm{d}t$ is a timelike interval, and therefore $t$ acts like a timelike coordinate; hence we refer to these spacetime regions as \textit{timelike} ($\mathrm{T}$) regions. 

Contrarily, when $B(r)<0$, $\mathrm{d}t$ is a spacelike interval, and hence we refer to such regions as \textit{spacelike} (S). An example of such a region is the interior of the event horizon $r=2\beta$ in the GRS metric~\eqref{eq:GRS}. Furthermore, if this S region is exterior to a cosmological horizon or a naked spacelike singularity, we classify it as S$^+$ due to the trajectories of particles being directed outwards ($\mathrm{d}r>0$); if the S region is interior to an event horizon, the opposite is true ($\mathrm{d}r<0$) so we classify it as S$^-$. 

Consequently, when we refer to a timelike or spacelike singularity, this refers to either $B(r=0)\rightarrow+\infty$ or $B(r=0)\rightarrow-\infty$ respectively.

\subsection{Horizons}\label{subsec:horizons} 

Horizons are at radii for which the blackening factor $B(r)=0$ for nonrotating sources~\cite{turner2020}, or more generally, $g_{rr}\rightarrow\pm\infty$ in~\eqref{eq:sph_symm_BH}~\cite{yulo2025}. In the MK metric, a maximum of three horizons are possible in the positive $r$ domain~\cite{turner2020}. 

In the interest of brevity in the following sections, we present here five categories of horizons and introduce shorthand notation to classify the causal structure of the MK metric: 
\begin{enumerate} 
    \item Event horizon ($\text{H}_\text{E}$): For increasing $r$, the spacetime regions are ordered as spacelike to timelike ($\mathrm{S}^-\rightarrow\text{H}_\text{E}\rightarrow \mathrm{T}$). An S$^-$ region interior to $\text{H}_\text{E}$ enforces inward trajectories towards $r=0$. 
    \item Cauchy horizon ($\text{H}_\text{C}$): For increasing $r$, $\mathrm{T}\rightarrow\text{H}_\text{C}\rightarrow \mathrm{S}^-(\rightarrow\mathrm{H}_\mathrm{E}\rightarrow \mathrm{T})$. In GR, this horizon is associated with charged or rotating black holes~\cite{moralesherrera2024}. $\text{H}_\text{C}$ only exists interior to $\text{H}_\text{E}$. 
    \item Cosmological horizon ($\text{H}_\Lambda$): $\mathrm{T}\rightarrow\text{H}_\Lambda\rightarrow \mathrm{S}^+$ for increasing $r$, like for $\text{H}_\text{C}$. $\text{H}_\Lambda$, when it exists, is always an outermost horizon~\cite{turner2020}, which distinguishes itself from $\mathrm{H}_\mathrm{C}$. The $\mathrm{S}^+$ region exterior to $\text{H}_\Lambda$ enforces outward trajectories to $r\rightarrow+\infty$~\cite{mitra2012}. 
    \item Extremal horizon ($\text{H}_\text{X}$): 
    $\mathrm{T}\rightarrow\text{H}_\text{X}\rightarrow \mathrm{T}$ for increasing $r$, which involves an exterior $\text{H}_\text{E}$ annihilating with an interior $\text{H}_\text{C}$. Such extremal horizons are associated with extremal metrics in GR~\cite{senjaya2025, horowitz2023}. 
   \item ``Nariai'' horizon ($\text{H}_\text{N}$):  $\mathrm{S}^-\rightarrow\text{H}_\text{N}\rightarrow \mathrm{S}^+$ for increasing $r$, involving an interior $\text{H}_\text{E}$ annihilating with an exterior $\text{H}_\Lambda$~\cite{castro2023}. These are often associated with the cosmological Nariai limit in GR metrics which contain the cosmological constant~\cite{anninos2010}.
\end{enumerate} 
For a more comprehensive definition of these types of horizons, consult~\cite{kusano2025_CGRN, yulo2025}. 

\subsection{Conditions for the existence of photon spheres}\label{subsec:ps_conditions} 

Null geodesics describe the trajectories of null particles in curved spacetime. From Noether's theorem and the Killing vectors $\xi_\mu$ which describe the symmetries of the metric, there exists a conserved quantity $E$ under an evolution in $t$, and a quantity $L$ over evolutions in $\phi$. Note however that in spacetimes that are not asymptotically flat, $E$ and $L$ do not directly represent energy and angular momentum~\cite{stuchlik2004}.

Using the geodesic equation describing trajectories in spacetime, we obtain the four equations of motion for null particles, as given below~\cite{turner2020, hoseini2017, stewart1989}: 
\begin{align} 
    \dfrac{\mathrm{d}t}{\mathrm{d}\lambda} &= \dfrac{E}{B(r)}, \label{eq:dt}\\
    \left(\dfrac{\mathrm{d}r}{\mathrm{d}\lambda}\right)^2&= E^2 - \dfrac{r^2 B(r)}{2} \left(\dfrac{\mathrm{d}\theta}{\mathrm{d}\lambda}\right)^2-\dfrac{L^2B(r)}{r^2\sin^2\theta}, \label{eq:dr}\\
     \dfrac{\mathrm{d}}{\mathrm{d}\lambda}\left(r^2 \dfrac{\mathrm{d}\theta}{\mathrm{d}\lambda}\right)&= \dfrac{L^2 \cos\theta}{r^2\sin^3\theta},\label{eq:dtheta}\\
    \dfrac{\mathrm{d}\phi}{\mathrm{d}\lambda} &= \dfrac{L}{r^2 \sin^2\theta}, \label{eq:dphi}   
\end{align}
where $\lambda$ is an affine parameter. Due to spherical symmetry, we may further confine our problem to the equatorial plane $\theta=\pi/2$ and combine~\eqref{eq:dr} and~\eqref{eq:dphi}. The dependence of $r$ on the azimuth $\phi$ is given by: 
\begin{align}\label{eq:vel} 
    \left(\dfrac{\mathrm{d}r/\mathrm{d}\lambda}{\mathrm{d}\phi/\mathrm{d}\lambda}\right)^2=\left(\dfrac{\mathrm{d}r}{\mathrm{d}\phi}\right)^2 
    = r^4\left(\left(\dfrac{E}{L}\right)^2-V_\text{eff}(r)\right) ,
\end{align} 
where the effective gravitational potential $V_\text{eff}(r)$ is:
\begin{equation}\label{eq:V_eff}
    V_\text{eff}(r)=\dfrac{B(r)}{r^2}=\dfrac{w}{r^2}-\dfrac{2M}{r^3} + \dfrac{\gamma}{r}-\kappa,
\end{equation}
the stationary points of which are given by:
\begin{align}
    {r_\mathrm{ust}}&=\dfrac{-w+1}{\gamma}=3\beta,\label{eq:rust}\\ 
    r_\mathrm{st}&=\dfrac{-w-1}{\gamma}=\dfrac{3\beta\gamma-2}{\gamma},\label{eq:rst}
\end{align}
with~\eqref{eq:rust} corresponding to the unstable maximum of~\eqref{eq:V_eff} and~\eqref{eq:rst} the stable minimum. 

\begin{figure}[t!]
    \centering
    \includegraphics[width=0.98\linewidth]{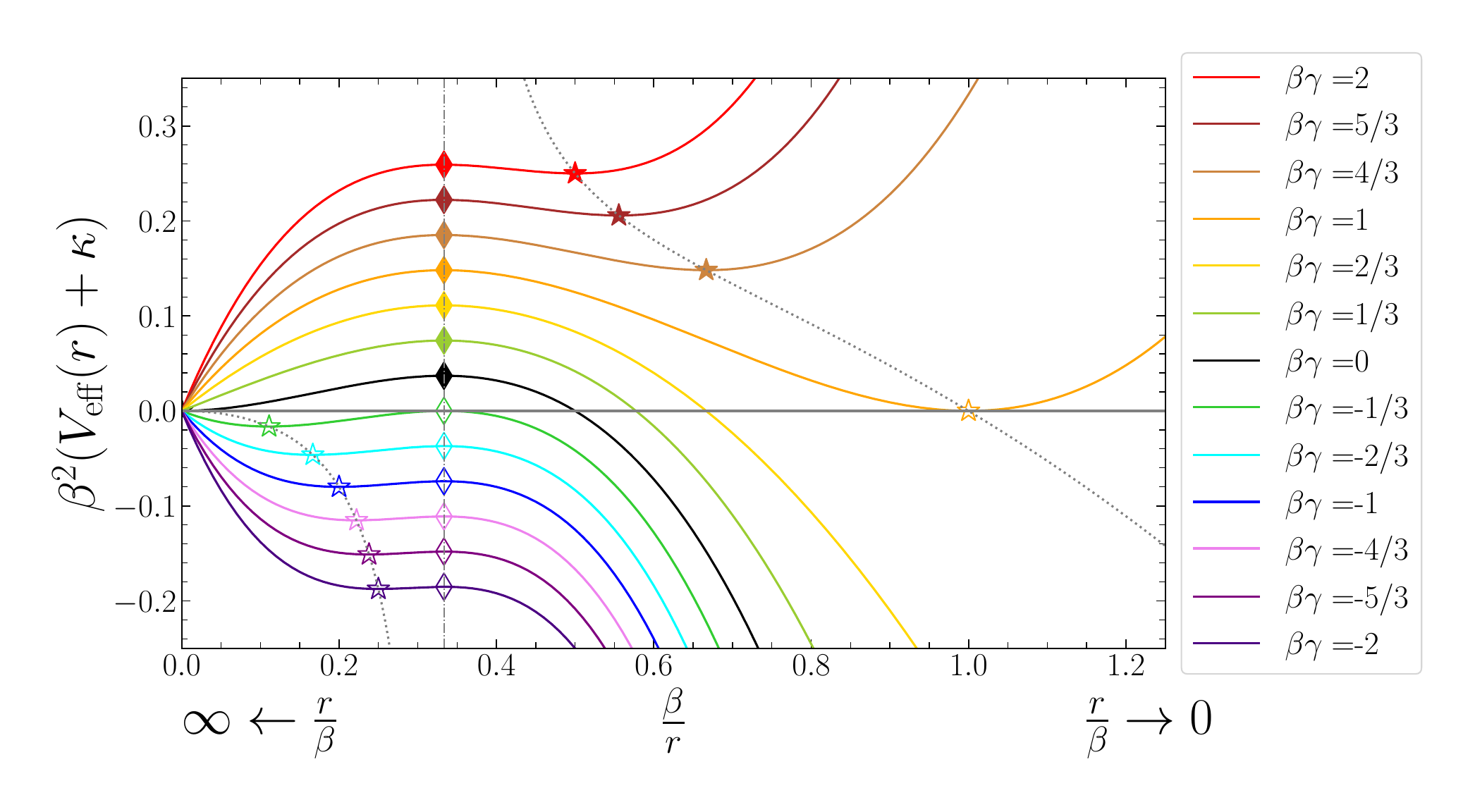}
    \caption{The effective potential $V_\text{eff}(r)$~\eqref{eq:V_eff} experienced by null particles in the MK metric~\eqref{eq:MK_metric} against $\beta/r$ for $-2\leq\beta\gamma\leq2$ in steps of $1/3$. $\beta/r$ as the radial parameter maps $r/\beta\rightarrow+\infty$ to $\beta/r=0$ and $r/\beta=0$ to $\beta/r\rightarrow+\infty$; as a result, the $V_\mathrm{eff}$ intercept corresponds naturally to the cosmological curvature $\kappa$. Stars/rhombuses denote stable/unstable photon spheres, and solid/hollow markers show whether they exist in $\mathrm{T}$/$\mathrm{S}$ regions. The dotted/dashed grey lines denote stable/unstable photon sphere radii at arbitrary $\beta\gamma$.} 
    \label{fig:V_eff_unperturbed} 
\end{figure} 

\begin{figure}[h!]
    \centering
    \includegraphics[width=0.98\linewidth]{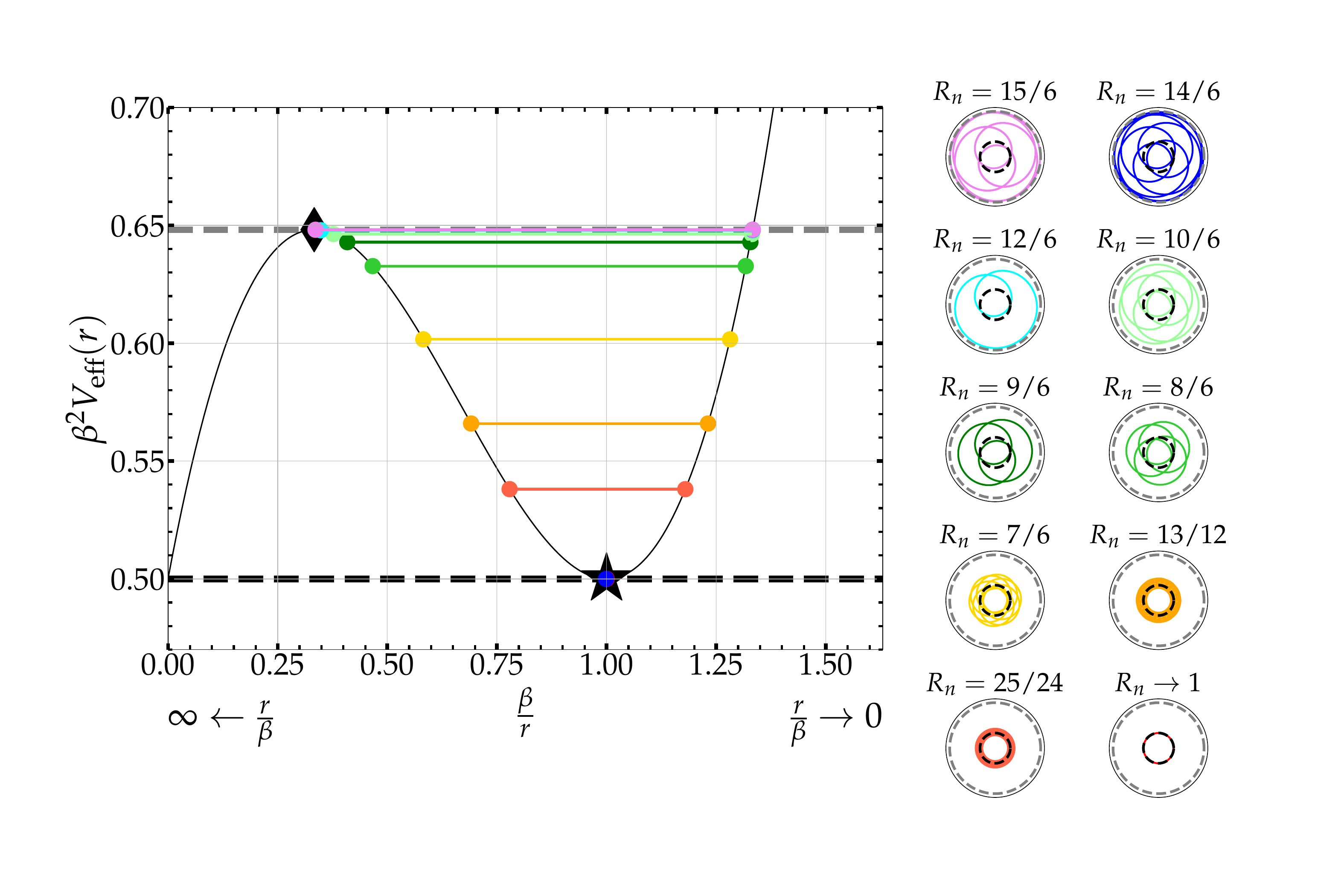}
    \caption{Energy levels of resonant null orbits for $V_\mathrm{eff}(r;\,\beta\gamma=1,\,\beta^2\kappa=-0.5)$ with their corresponding closed rosette orbits visualised to the right. The grey dashed lines correspond to $V_\mathrm{eff}(r=r_\mathrm{ust})$, and the black dashed lines correspond to $V_\mathrm{eff}(r=r_\mathrm{st})$. $R_n$ corresponds to the ratio of azimuthal oscillations to radial oscillations.} 
    \label{fig:rosettes}
\end{figure}

\begin{figure}
    \centering
    \includegraphics[width=0.72\linewidth]{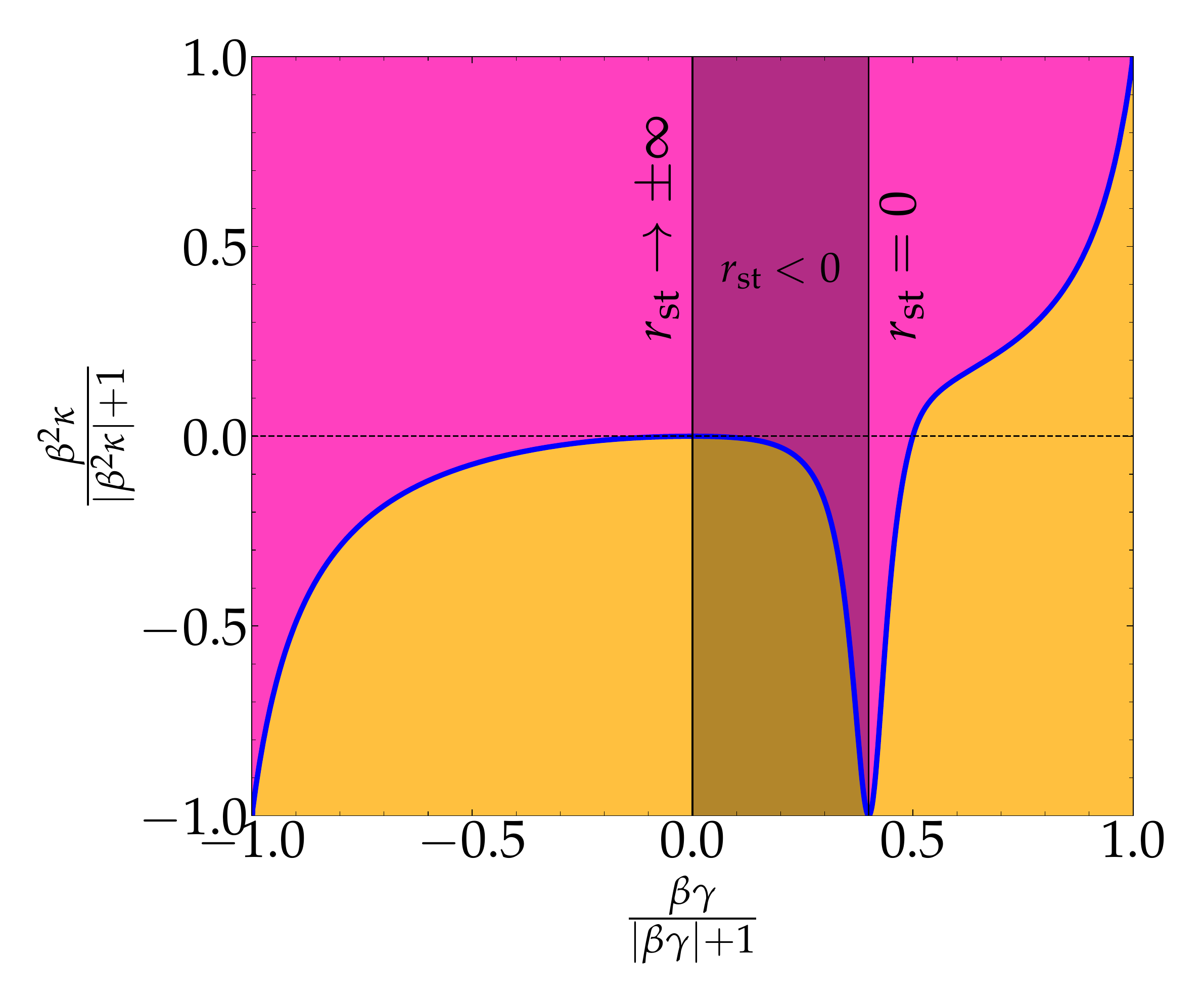}
    \caption{A dimensionless parameter map spanning $(-\infty, +\infty)$ in both $\beta\gamma$ and $\beta^2\kappa$, showing whether $r_\mathrm{st}$ exists in a $\mathrm{T}$ (yellow) or $\mathrm{S}$ (pink) region. The blue boundary corresponds to~\eqref{eq:ps_hor}. The darkened region corresponds to $r_\mathrm{st}<0$.} 
    \label{fig:turner_map_inf} 
\end{figure}

$V_\mathrm{eff}$ describes the trajectories of null particles under the effect of the central gravitational source, and is illustrated in figure~\ref{fig:V_eff_unperturbed} for different $\beta\gamma$ values. The stationary points of $V_\text{eff}$ correspond to the photon spheres of the metric, with unstable local maxima at $r_\text{ust}=3\beta$ shown by rhombuses and stable local minima at $r_\mathrm{st}=3\beta-2/\gamma$ shown by stars. As expected, photon orbits near $r_\mathrm{st}$ can give rise to resonant orbits as seen in figure~\ref{fig:rosettes}, due to the stability of the potential minimum. These photons are bounded and trapped within the potential well.

However, for photon orbits to be physically feasible, we require additional considerations. First, as $\beta>0$ in the present work, $0\leq\beta\gamma\leq2/3$ yields negative or undefined values of $r_\mathrm{st}$. Therefore, $\beta\gamma > 2/3$ or $\beta\gamma < 0$ for $r_\mathrm{st}>0$ in positive-$\beta$ spacetimes. 

Moreover, even if $r_\mathrm{st}>0$, this still does not guarantee the existence of a stable photon sphere: a photon sphere can only exist in $\mathrm{T}$ regions where $B(r_\mathrm{st})>0$. $\mathrm{S}$ regions with $B(r_\mathrm{st})<0$ enforce trajectories of increasing or decreasing $r$, meaning that circular orbits of constant $r$ are inherently forbidden. This distinction is shown on figure~\ref{fig:V_eff_unperturbed} by solid markers for physically meaningful photon spheres in $\mathrm{T}$ regions, and hollow markers for unphysical ones in $\mathrm{S}$ regions. From $B(r_\mathrm{st})=0$ and~\eqref{eq:MK_metric}, the stable photon sphere collides with a horizon when~\cite{turner2020}: 
\begin{equation}\label{eq:ps_hor}
    \beta^2\kappa = \dfrac{\beta\gamma - 1}{(3 - 2/\beta\gamma)^2},
\end{equation}
such that physically feasible stable photon spheres exist for: 
\begin{equation}\label{eq:ps_cond}
    \beta^2\kappa < \dfrac{\beta\gamma - 1}{(3 - 2/\beta\gamma)^2}. 
\end{equation} 
A map of these parameters is shown in figure~\ref{fig:turner_map_inf}, serving as a guideline to selecting appropriate values of $\beta\gamma$ and $\beta^2 \kappa$, such that physically meaningful stable photon sphere radii may be obtained. We exclusively select from the yellow domain, corresponding to $\mathrm{T}$ regions. Except for $\beta\gamma\rightarrow-\infty$ and $\beta\gamma=2/3$, there is always a 
negative-enough value of $\beta^2\kappa$ that will allow a photon sphere to exist in a T region; crucially, while $\beta^2\kappa$ does not affect the radius of the photon sphere, it affects the sign of $B(r_\mathrm{st})$ by shifting the potential $V_\mathrm{eff}$ up or down, as seen in~\eqref{eq:V_eff}. The region where $r_\mathrm{st}\leq0$ has been blocked out by the dark rectangle; in this region, $r_\mathrm{st}>0$ if $\beta<0$. Again, we do not consider such solutions which correspond to negative mass. 

This aligns with our expression for $B(r)$ and can be understood with reference to~\eqref{eq:MK_metric} and~\eqref{eq:V_eff}; $\beta^2\kappa<0$ corresponds to a positive contribution from the quadratic term to the blackening factor $B(r)$, which means that there is a higher likelihood for $B(r_\mathrm{st})$ to be in a $\mathrm{T}$ region. Notably, $\beta^2\kappa\geq0$ is forbidden for circular photon orbits to exist in $\beta\gamma<0$ spacetimes. 

An analysis of the topology of stationary black holes in asymptotically flat spacetimes in both GR and modified gravity theories has shown the guaranteed presence of at least one photon sphere radius in such spacetimes~\cite{AsymptoticFlatness}. While the metric we are considering here is not asymptotically flat due to the long-range linear and quadratic terms, the aforementioned work does not preclude the existence of photon spheres in non-asymptotically flat spacetimes. For instance, stationary GR spacetimes with de Sitter and Anti-de Sitter backgrounds still possess photon spheres~\cite{KerrdS, KerrAdS}.

\section{Satisfying the field equations}\label{sec:geometries_pert} 

\subsection{Thin shell sources}\label{subsec:pressure}

We now partition our sourceless MK spacetime into two regions: an interior region $\mathcal{V}^{-}$ that contains a singular source described by a metric tensor $g_{\mu\nu}^-$, and an exterior region $\mathcal{V}^+$ described by $g_{\mu\nu}^{+}$~\cite{poisson2004}. These two solutions are segregated by a hypersurface of width zero, corresponding to our shell of photons or null particles; this then means that $f(r)\neq0$. Through this, we alter the spacetime metric by loading the photon sphere at $r_\mathrm{st}$ with a thin shell cavity, which surrounds an unresolved source with $B(r=0)\rightarrow\pm\infty$. Consequently, the MK blackening factor is of the form: 
\begin{equation}\label{eq:MK_perturbed} 
    B(r) = w(r)-\dfrac{2M(r)}{r}+\gamma(r) r -\kappa(r) r^2.
\end{equation} 
These new $r$-dependent MK parameters must then further fulfill the third-order constraint of~\eqref{eq:3_order} at every radius. By using appropriate boundary conditions, the MK parameters from~\eqref{eq:MK_perturbed} are readily obtained from integrating first-order equations arising from~\eqref{eq:Poisson_4}~\cite{mannheim2007_2, horne2016, horne2019}: 
\begin{equation}\label{eq:moments_w_boundary} 
    \begin{aligned} 
        w'(r) &= \frac{+r^3}{2} f(r), & \quad w(r=0) &= w_0 , \\
        M'(r) &= \frac{+r^4}{12} f(r), & \quad M(r=0) &= M_0 , \\
        \gamma'(r) &= \frac{-r^2}{2} f(r), & \quad \gamma(r=0) &= \gamma_0, \\
        \kappa'(r) &= \frac{-r}{6} f(r), & \quad \kappa(r\rightarrow+\infty) &= \kappa_\infty,
    \end{aligned}
\end{equation} 
with the auxiliary parameters $w_0$ and $M_0$ being given as 
\begin{equation}\label{eq:w&M_vacuum}
    \begin{aligned}
        w_0 &= 1 - 3\beta_0\gamma_0,\\
        M_0 &=\beta_0 \left(1-\dfrac{3}{2}\beta_0\gamma_0\right),
    \end{aligned}
\end{equation}
if the interior solution is a vacuum. Here, $\beta_0$, $\gamma_0$, and $\kappa_\infty$ are the constant parameters originally existing in the sourceless blackening factor $B_0(r)$ in~\eqref{eq:MK_metric} with $f(r)=0$. $M(r)$ and $\gamma(r)$ have boundary conditions at $r=0$ due to the presence of the aforementioned unresolved singular structure at the origin, while in contrast $\kappa(r)$ is anchored at $r\rightarrow+\infty$ due to its aforementioned connection to the cosmological curvature~\cite{mannheim2007_2}. For $w(r)$, we have the freedom to fulfill~\eqref{eq:3_order} either inside or outside the thin shell, if the radial pressure ${T^{r}}_{r}=p_r$ is known at either of these locations~\cite{horne2016}. Therefore, to patch two spacetimes $g^-_{\mu \nu}$ and $g^+_{\mu \nu}$ together in CG in the way that we are doing in the present work, we require an adjustment of both the interior ($w(r)$,~$M(r)$,~$\gamma(r)$) and exterior ($\kappa(r)$) CG parameters as a function of the strength of the load. 

With the equations presented in~\eqref{eq:moments_w_boundary} and the shell placed at an appropriate radius, we ensure that the union of the two regimes interior and exterior to the photon shell together form a valid static solution to the Bach field equations~\eqref{eq:BE}. As our hypersurface naturally has zero radial width, we must assure that ${T^r}_r=0$ immediately interior to and exterior to the shell. From this, since the third-order constraint~\eqref{eq:BE_rr} must necessarily be fulfilled at every radius in $B(r)$, we choose to set the initial condition of $w$ inside of the shell.

We first investigate the consequences of an infinitely-thin shell of arbitrary radius $r_\mathrm{sh}$. This can be described by a Dirac $\delta$ function as: 
\begin{equation}\label{eq:fr_dirac} 
    f(r)=f_0\,\delta(r-r_\mathrm{sh}) ,
\end{equation} 
where $f_0 (\geq 0)$ is defined from the third-order constraint~\eqref{eq:3_order}: 
\begin{equation}\label{eq:f0}
    f_0 = \dfrac{-3}{4\alpha_\mathrm{g}}\int \mathrm{d}r \dfrac{\rho(r) + p_r(r)}{B(r)}. 
\end{equation}
$f_0$ is an arbitrary loading amplitude describing the density of the null matter distribution in the shell at $r=r_\mathrm{sh}$; we can turn $f_0$ up and down to control the density of null matter within the thin shell, the degree to which the two regions of spacetime are discontinuous from one another. While not strictly being a physically feasible distribution of null particles, we employ this as a toy model: if we presume that the majority of the orbiting null matter is orbiting at radii very close to $r=r_\mathrm{sh}$, this is justifiable. Furthermore, while strictly speaking the Dirac $\delta$ distribution would be instead associated with $\rho$ and $p_r$ individually, we simplify our problem by amalgamating the infinite contributions from both $\rho$ and $p_r$ into $f(r)$, which describes more or less the sum of these two components of the energy-momentum tensor. This formulation gives us physically intuitive results and easily adjustable parameters. 

Using~\eqref{eq:moments_w_boundary}, we find: 
\begin{equation}\label{eq:MK_funcs}
\begin{aligned}
    w(r) &= w_0+\Delta w \hspace{0.1cm}\Theta(r-r_\mathrm{sh}), &\quad \Delta w &= +\dfrac{r_\mathrm{sh}^3}{2}f_0,\\
    M(r) &= M_0+\Delta M \hspace{0.1cm}\Theta(r-r_\mathrm{sh}), &\quad \Delta M &= +\dfrac{r_\mathrm{sh}^4}{12}f_0,\\
    \gamma(r) &= \gamma_0+ \Delta \gamma \hspace{0.1cm}\Theta(r-r_\mathrm{sh}), &\quad \Delta \gamma &= -\dfrac{r_\mathrm{sh}^2}{2}f_0,\\
    \kappa(r) &= \kappa_\infty+\Delta\kappa\hspace{0.1cm}\left(\Theta(r-r_\mathrm{sh}) - 1\right), &\quad \Delta\kappa &= -\dfrac{r_\mathrm{sh}}{6}f_0,
\end{aligned}
\end{equation}
where the $\Delta$ parameters are the jumps in the MK parameters across the photon shell, and $\Theta(x-x_0)$ is the Heaviside $\Theta$ function defined by: 
\begin{equation}
\Theta(x-x_0)=
    \begin{cases}
        0 & \text{if } x < x_0, \\
        1 & \text{if } x \geq x_0.
    \end{cases}
\end{equation}
As the MK parameters~\eqref{eq:MK_funcs} are step functions that jump at $r=r_\mathrm{sh}$, the metric $g_{\mu\nu}^{-}$ of the interior spacetime solution $\mathcal{V}^-$ is naturally associated with the blackening factor
\begin{equation}
    B_-(r)=w_0 - \dfrac{2M_0}{r}+\gamma_0 r - \left(\kappa_\infty -\Delta \kappa\right)r^2,
\end{equation}
while the exterior solution $\mathcal{V}^+$ must have
\begin{equation}
    B_+(r)=\left(w_0 + \Delta w\right)- \dfrac{2\left(M_0 + \Delta M\right)}{r} + \left(\gamma_0 + \Delta \gamma\right)r - \kappa_\infty r^2,
\end{equation}
in its metric tensor $g_{\mu\nu}^+$.

\subsection{Pressure jump across the shell}\label{subsec:pressure_jump}

\begin{figure}
    \centering
    \includegraphics[width=0.98\linewidth]{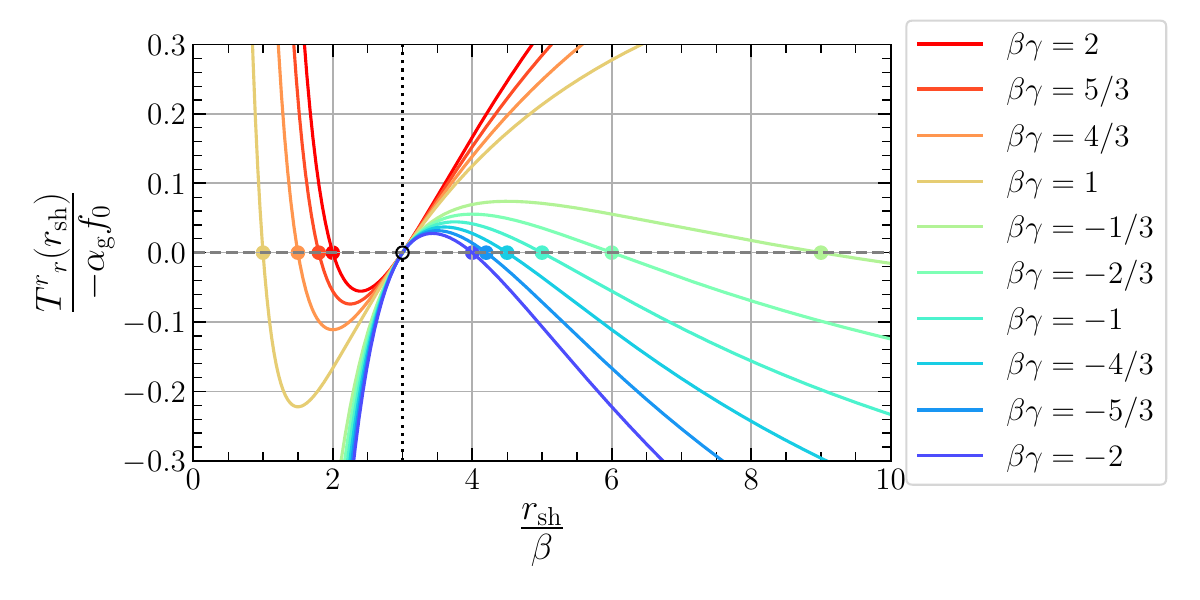}
    \caption{The jump in pressure~\eqref{eq:pressure_jump} for different values of $\beta\gamma$, arising as a result of the infinitely thin shell of an arbitrary radius $r_\mathrm{sh}/\beta$. Predictably,~\eqref{eq:pressure_jump} has two roots: one at the unstable photon sphere $r_\mathrm{ust}$ (hollow black circle), and the stable photon sphere $r_\mathrm{st}$ (coloured filled circles).} 
    \label{fig:pressure_jump}
\end{figure}

Additional considerations must be made to guarantee fulfillment of the CG field equations: namely, whether ${T^r}_r=p_r(r<r_\text{sh})=p_r(r>r_\text{sh})$ for any arbitrary shell radius $r_\text{sh}$. We assume for simplicity that the interior solution is a vacuum with $T_{\mu\nu}=0$; of course while this is not a necessity, and in fact the radial pressure can be nonzero on either side, for the solution to be static we require ${T^r}_r$ to not jump across the shell. Setting ${T^r}_r(r<r_\mathrm{sh})=0$ is a convenient way to approach this. From this, we can enforce ${T^r}_r(r>r_\mathrm{sh})=p_r(r>r_\mathrm{sh})=0$.~\eqref{eq:3_order} becomes: 
\begin{equation}\label{eq:3_order_jump_req}
\begin{aligned}
    0 =& \left(w_0+\Delta w\right)^2 - \left(1 - 6\left(M_0+\Delta M\right)\left(\gamma_0+\Delta \gamma\right) \right).
\end{aligned}
\end{equation} 
From the third order constraint inside the shell, $w_0^2 + 6M_0\gamma_0 - 1=0$. Also, from~\eqref{eq:MK_funcs}:
\begin{equation}
    \Delta w^2 + 6\Delta M\Delta \gamma=r_\mathrm{st}^6 f_0^2\left(\dfrac{1}{4}+6\left(\dfrac{1}{12}\cdot\left(-\dfrac{1}{2}\right)\right)\right)=0.
\end{equation}
As a result,~\eqref{eq:3_order_jump_req} simplifies to: 
\begin{equation}
\begin{aligned}
    2w_0 \Delta w + 6M_0\Delta \gamma + 6\Delta M\gamma_0=
    \dfrac{r_\text{sh}^2 f_0}{2}\left(\gamma_0 r_\text{sh}^2 + 2w_0r_\text{sh} - 6M_0\right) 
    =0.
\end{aligned}
\end{equation}
For the above to hold, we solve the quadratic by applying the third-order constraint~\eqref{eq:3_order}: 
\begin{align}\label{eq:r_sh}
    r_\text{sh}&=\dfrac{-w_0\pm\sqrt{w_0^2 + 6M_0\gamma_0}}{\gamma_0} = \dfrac{-w_0\pm 1}{\gamma_0}, 
\end{align}
yielding exactly the same roots as the photon sphere radii~\eqref{eq:rust} and~\eqref{eq:rst} in the unloaded ($f_0=0$) metric. This is a key result: outside an infinitely thin shell of radius $r_\text{sh}$ that is defined by a Dirac $\delta$ distribution in $f(r)$, the difference in the radial pressure inside and outside the shell is zero \textit{if and only if} $r_\text{sh}=r_\text{ust}$ or $r_\text{sh}=r_\mathrm{st}$; a shell of any other radius induces a discontinuity in pressure. The magnitude of this pressure jump may be given by rearranging the third-order constraint~\eqref{eq:3_order}:
\begin{equation}\label{eq:pressure_jump}
    \dfrac{{T^r}_r}{-\alpha_\mathrm{g} f_0} = \dfrac{2\gamma}{3 r_\mathrm{sh}}\left(r_\mathrm{sh}-r_\mathrm{st}\right)\left(r_\mathrm{sh}-r_\mathrm{ust}\right),
\end{equation}
which is graphed in figure~\ref{fig:pressure_jump}. As expected, the left-hand side of~\eqref{eq:pressure_jump} reduces to zero at $r=r_\mathrm{ust}$ (black circle) or $r=r_\mathrm{st}$ (coloured circles). 

\subsection{Photon sphere radii}\label{subsec:PS-pert} 

As the MK parameters experience jumps from one side of the shell to the other, the radius of the shell may change from its original radius. As previously established, $V_\text{eff}(r)=B(r)/r^2$, so from~\eqref{eq:MK_perturbed}:
\begin{equation}
    V_\text{eff}(r)=\dfrac{w(r)}{r^2}-\dfrac{2M(r)}{r^3} + \dfrac{\gamma(r)}{r}-\kappa(r). 
\end{equation}
From~\eqref{eq:r_sh}, we can find a general formula for photon sphere radii:
\begin{equation}\label{eq:r_st_perturb}
    r_{\mathrm{(u)st}} = \dfrac{-(w_0+\Delta w)\mp1}{\gamma_0+\Delta\gamma},
\end{equation}
where the negative root corresponds to $r_{\mathrm{st}}$ and the positive root to $r_{\text{ust}}$. For photon sphere radii evaluated outside the photon shell one can substitute the values for $\Delta w$ and $\Delta \gamma$ from~\eqref{eq:MK_funcs}, while the photon sphere radii as evaluated by the interior MK parameters may be recovered by setting $\Delta w$ and $\Delta \gamma$ to 0. 

To find whether loading this thin shell actually makes a difference in the radius of the stable photon sphere, we recall~\eqref{eq:MK_funcs},~\eqref{eq:r_sh}, and~\eqref{eq:r_st_perturb}. Setting $r_\mathrm{st}^-$ and $r_\mathrm{st}^+$ to be the stable photon sphere radius evaluated inside and outside the shell respectively, the jump in the (stable) photon sphere radius $\Delta r$ may be evaluated to give: 
\begin{equation}
\begin{aligned}
    \Delta r = r_\mathrm{st}^- - r_\mathrm{st}^+&= \dfrac{-w_0-1}{\gamma_0}-\left(-\dfrac{\left(w_0+ \dfrac{(r_\mathrm{st}^-)^3f_0}{2}+1\right)}{\gamma_0-\dfrac{(r_\mathrm{st}^-)^2f_0}{2}}\right)\\
    &=\dfrac{(r_\mathrm{st}^-)^2f_0}{2}\dfrac{w_0+1+r_\mathrm{st}^-\gamma_0}{\gamma_0\left(\gamma_0-\dfrac{(r_\mathrm{st}^-)^2f_0}{2}\right)}.
\end{aligned}
\end{equation}

Remembering that $r_{\mathrm{st}}^{-}\gamma_0=-w_0-1$ from~\eqref{eq:r_sh}, $r_{\mathrm{st}}^{-}=r_{\mathrm{st}}^{+}$ and therefore the jump in photon sphere radius $\Delta r=0$. This is also a major result: while loading the stable photon sphere changes the overall structure of spacetime enough to move an exterior $r_\text{ust}$, this Dirac $\delta$ function actually has \textit{no effect} on the area of the photon sphere at which the null matter is accumulating. Linking this to the result from earlier regarding the third-order constraint, there seems to be a significance associated with these photon sphere radii of CG, which extends beyond just their established definition as the radii of circular null geodesics~\cite{mannheim2025_perscomm}. 

We note several discrepancies between our CG results, versus those found in similar studies conducted in GR. 

First, we note that unlike what is found in~\cite{diFilippo2025}, the accumulation of photons at the stable photon sphere of the MK metric does not give rise to additional spherical orbits. In our case, this may be justified by a simple algebraic argument: the interior ($r<r_\mathrm{st}$) and exterior ($r>r_\mathrm{st}$) effective potentials $V_\mathrm{eff}$ generated by the MK metric together form a smooth cubic function in $\beta/r$, which at most have one unstable maximum and one stable minimum. The unstable maximum will not move when $r_\mathrm{ust}<r_\mathrm{st}$ and will move if $r_\mathrm{ust}>r_\mathrm{st}$, but this in principle has no effect on the total number of photon spheres obtainable in the metric. 

Second,~\cite{cunha2025} finds that accumulating photons at the stable photon sphere of an ultracompact continuum-shell star in GR serve to push the photon sphere towards smaller $r$. Again, this may be justified by a purely algebraic argument: the presence of a linear $\gamma\, r$ in the blackening factor gives rise to the stable photon sphere radius $r_\mathrm{st}=(-1-w)/\gamma$. $w$ jumps up across the shell but $\gamma$ drops down, and the total change in $r_\mathrm{st}$ amounts to zero; in the absence of $\gamma$, such as in e.g. GRS, the sole contribution is from $w$, and the overall effect is to push the photon sphere inwards. Therefore, the contrast in the effect of photon accumulation between~\cite{cunha2025} and the present work may be an indication that CG spacetimes are generally less sensitive to perturbations and less prone to instability than GR spacetimes.

\section{Spacetime structures and horizons}\label{subsec:horizons-pert}

\begin{figure}[h!] 
    \centering 
    \begin{subfigure}[b]{0.49\textwidth}
        \includegraphics[width=\textwidth]{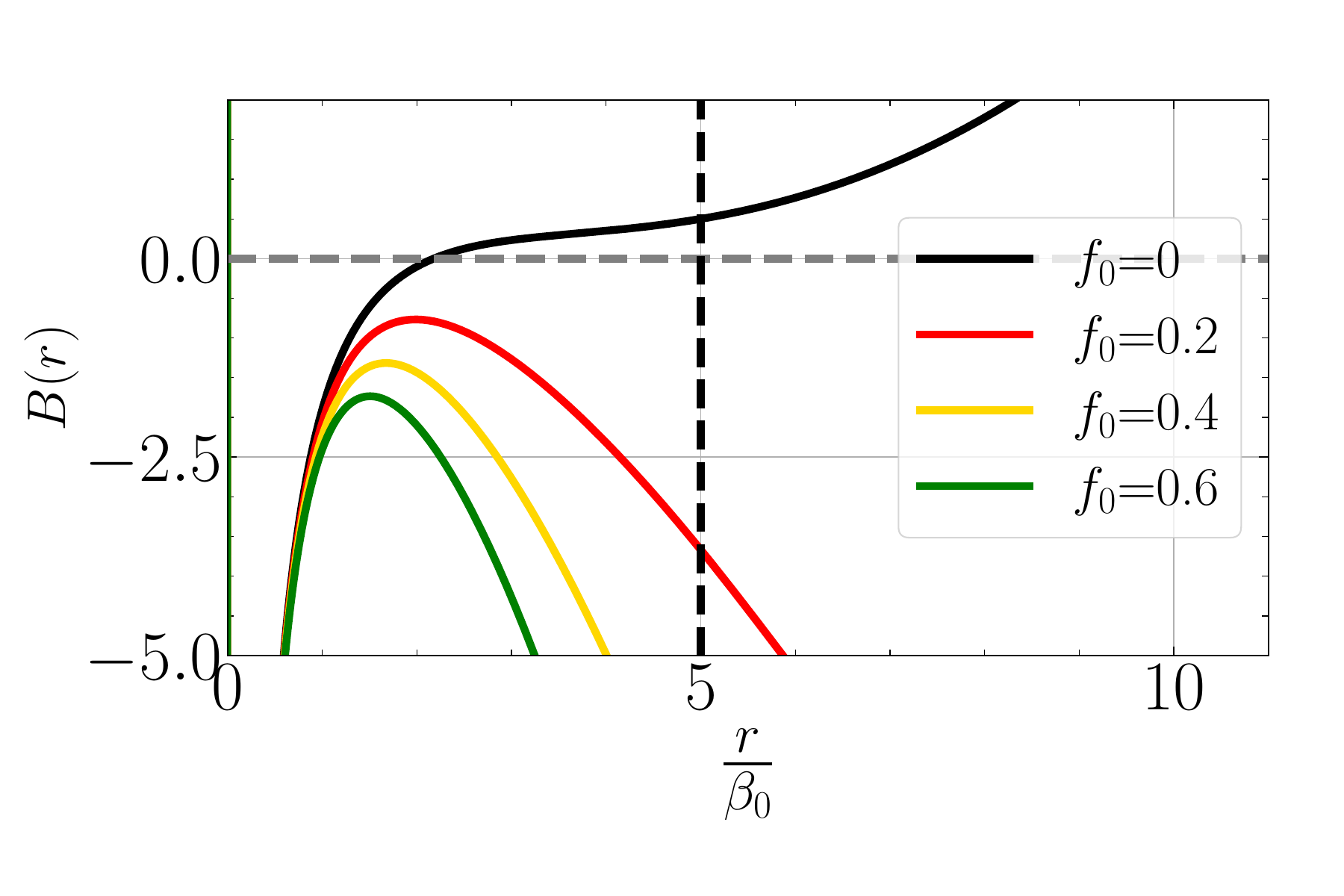}
        \caption[]
        {$\beta_0\gamma_0=-1, \beta_0^2\kappa_\infty=-0.1$.} 
        \label{subfig:B(r)_neg} 
    \end{subfigure} 
    \hfill
    \begin{subfigure}[b]{0.49\textwidth}  
        \includegraphics[width=\textwidth]{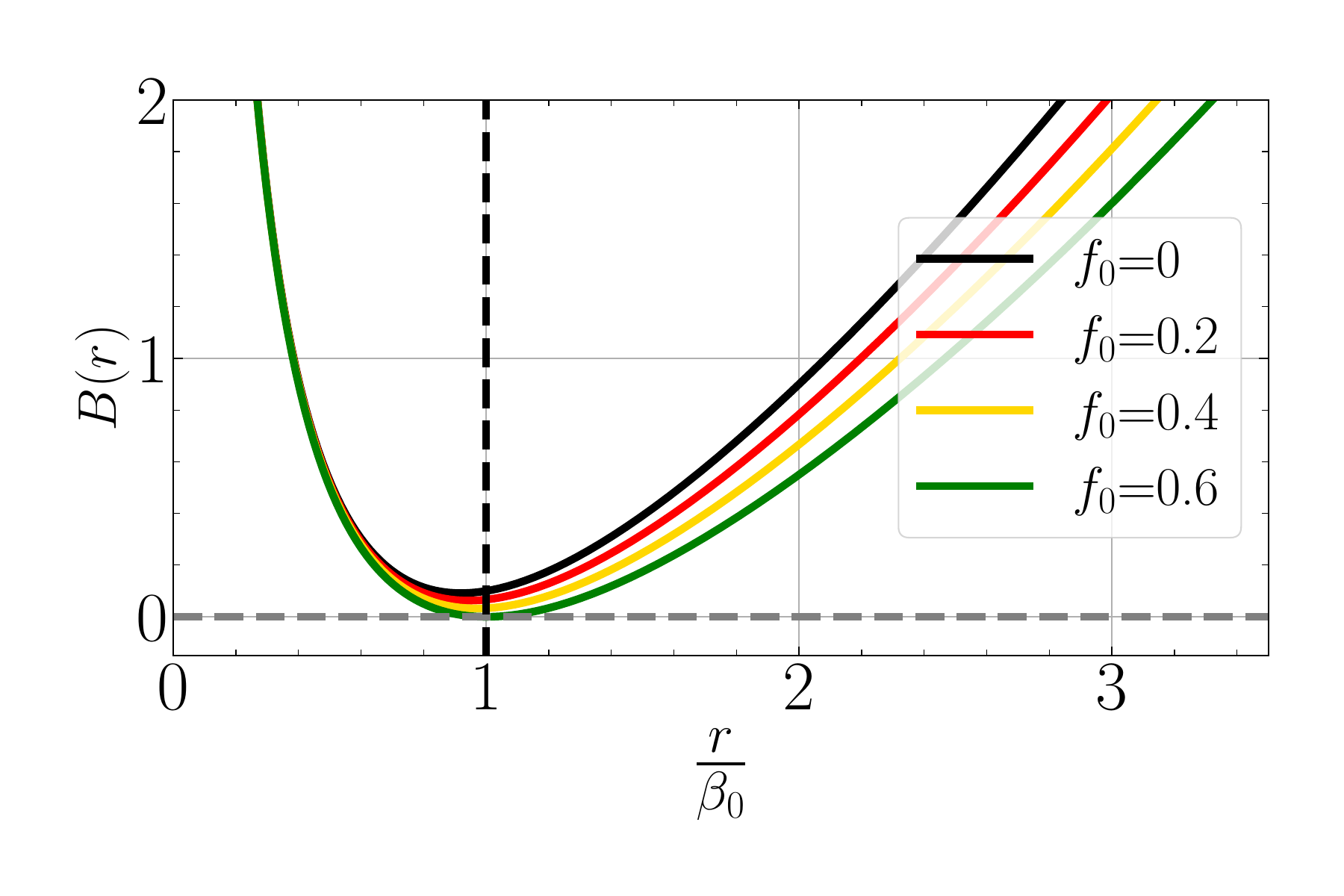}
        \caption[]%
        {$\beta_0\gamma_0=+1, \beta_0^2\kappa_\infty=-0.1$.}    
        \label{subfig:B(r)_pos}
    \end{subfigure}
    \caption[]
    {Plots of $B(r)$ for different configurations of initial MK parameter values and loading amplitudes. \textit{Dashed horizontal grey line}: $B(r)=0$, \textit{dashed vertical black line}: $r=r_\mathrm{st}$.} 
    \label{fig:B(r)_line}
\end{figure}

Upon loading the photon sphere, $r_\mathrm{st}$ does not change; however, the causal structure of the spacetime may indeed change. Plots of $B(r)$ in figure~\ref{fig:B(r)_line} show that $B(r)$ decreases as $f_0$ increases, regardless of whether the central singularity is spacelike as $B(r=0)\rightarrow-\infty$ (figure~\ref{subfig:B(r)_neg}) or timelike with $B(r=0)\rightarrow+\infty$ (figure~\ref{subfig:B(r)_pos}). Notably, there are no discontinuities in the blackening factor $B(r)$ despite the presence of the shell. 

$B(r)$ may then be represented for $-\infty<\beta_0\gamma_0<+\infty$ and $0\leq r/\beta_0<+\infty$, to probe the evolution of causal structures when the loading amplitude is increased. In the unloaded MK blackening factor $B_0(r)$, loci of horizons are given by the solutions of the cubic equation $\Delta^\text{H}$ below~\cite{turner2020}: 
\begin{equation}\label{eq:horiz_unperturbed}
    \Delta^\text{H}\equiv\left(\dfrac{r}{\beta_0}\right)B_0(r)=-\beta_0^2\kappa_\infty \left(\dfrac{r}{\beta_0}\right)^3 + \beta_0\gamma_0 \left(\dfrac{r}{\beta_0}\right)^2 + w_0 \left(\dfrac{r}{\beta_0}\right) - \dfrac{2M_0} {\beta_0}=0.
\end{equation} 

\begin{figure}[t!]
    \centering
    \begin{subfigure}[b]{0.999999\textwidth}
        \includegraphics[width=\textwidth]{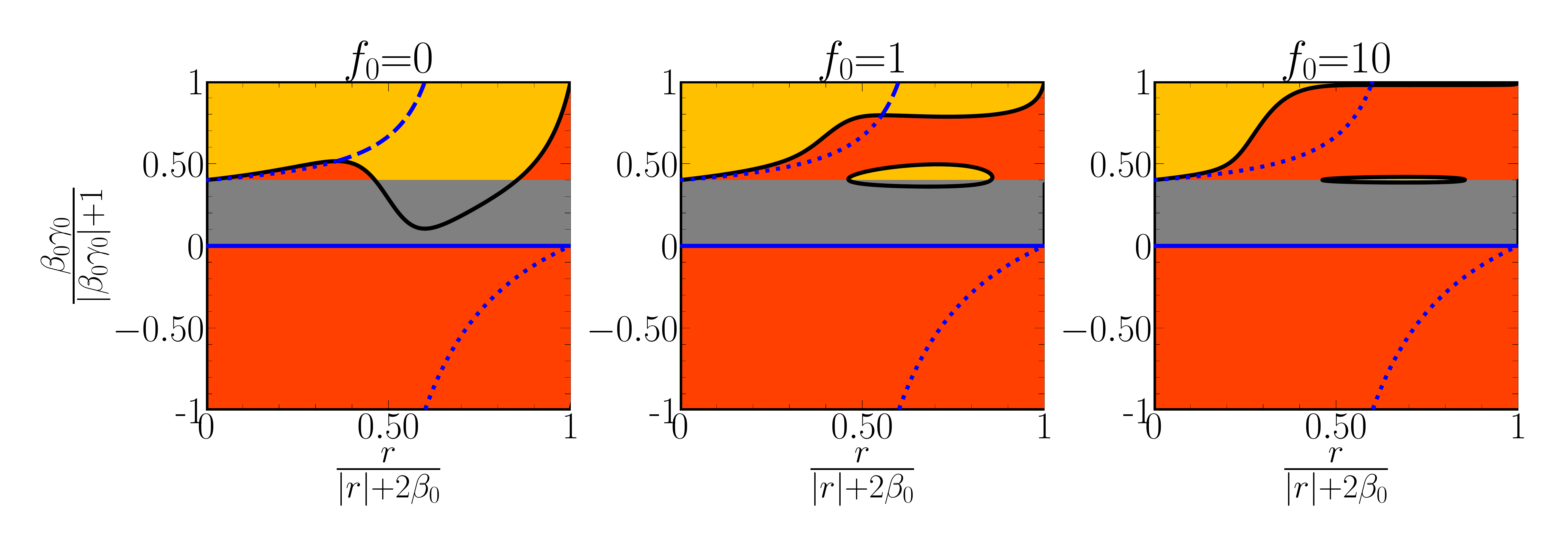}
        \caption[]
        {$\beta_0^2\kappa_\infty=+0.05$ (GR equivalent: GRSdS).}
        \label{subfig:horz_k>0}
    \end{subfigure}
    \hfill
    \begin{subfigure}[b]{0.999999\textwidth}
        \includegraphics[width=\textwidth]{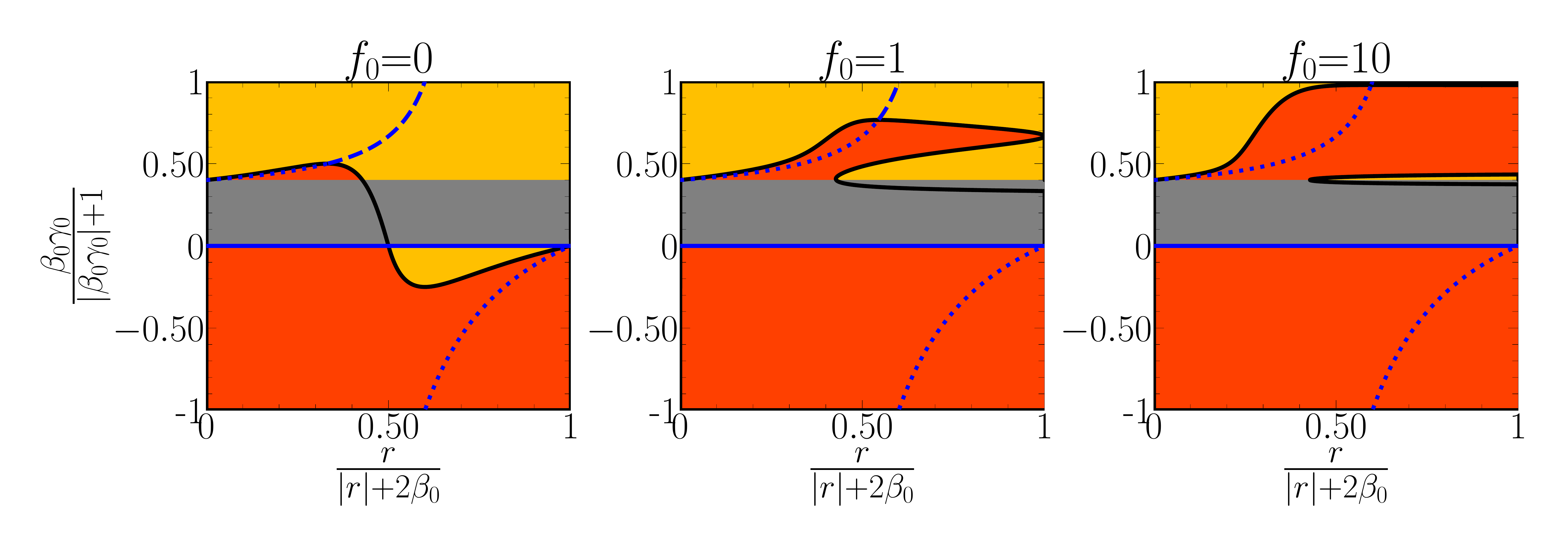}
        \caption[]
        {$\beta_0^2\kappa_\infty=0$ (GR equivalent: GRS).}
        \label{subfig:horz_k=0}
    \end{subfigure}
    \hfill
    \begin{subfigure}[b]{0.999999\textwidth}
        \includegraphics[width=\textwidth]{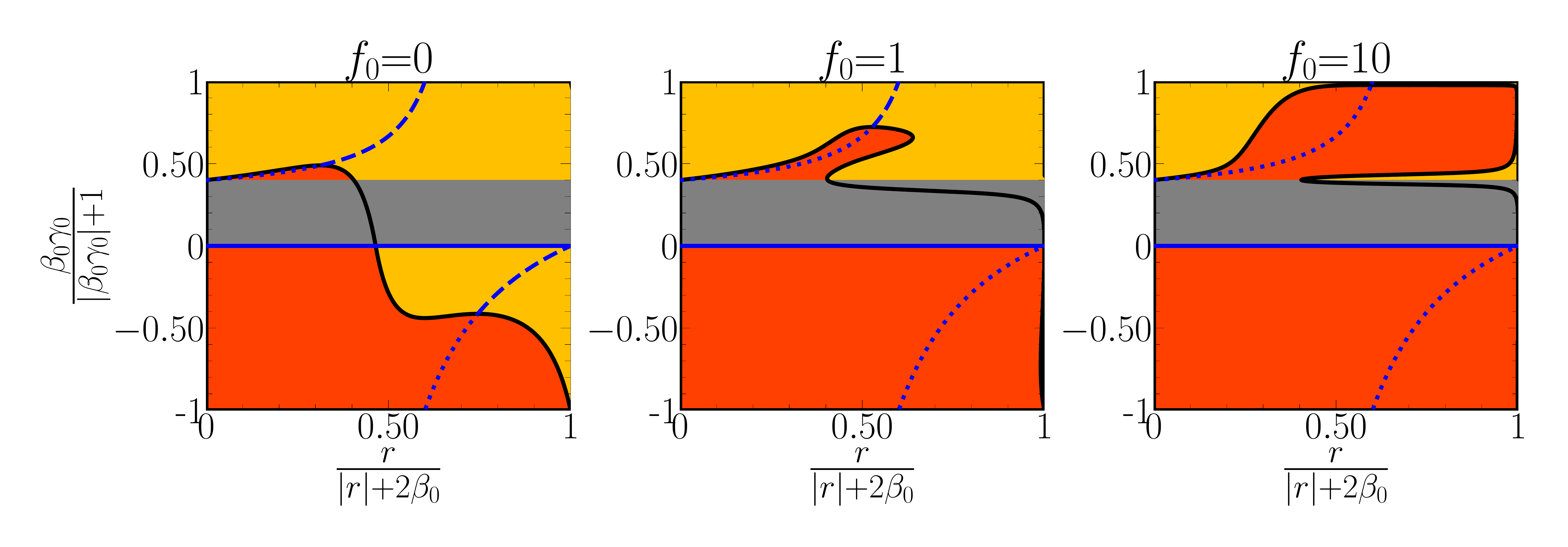}
        \caption[]
        {$\beta_0^2\kappa_\infty=-0.05$ (GR equivalent: GRSAdS).}
        \label{subfig:horz_k<0}
    \end{subfigure}
    \caption[]
    {Horizon plots for various values of $\beta_0^2\kappa_\infty$, for $-\infty<\beta_0\gamma_0<+\infty$ and $0\leq r/\beta_0<+\infty$. \textit{Yellow}:
    $\mathrm{T}$ regions, \textit{red}: $\mathrm{S}$ regions, \textit{solid black line}: horizons $\widetilde{\Delta}^\mathrm{H}=0$~\eqref{eq:horiz_perturbed}, \textit{blue dashed/dotted line}: $r=r_\mathrm{st}$~\eqref{eq:rst} in T/S regions, \textit{blue solid line}: equivalent metric in GR where $\beta_0\gamma_0=0$, \textit{grey regions}: $r_\mathrm{st}<0$. $f_0=0$ in each subfigure naturally corresponds to the unloaded metric.} 
    \label{fig:horizons} 
\end{figure} 

The horizons of the MK metric with a nonzero loading value $f_0$ are therefore the roots of $\widetilde{\Delta}^\text{H}$, where: 
\begin{equation}\label{eq:horiz_perturbed}
    \widetilde{\Delta}^\text{H}\equiv\left(\dfrac{r}{\beta_0}\right)B(r)=-\beta_0^2\kappa(r) \left(\dfrac{r}{\beta_0}\right)^3 + \beta_0\gamma(r) \left(\dfrac{r}{\beta_0}\right)^2 + w(r) \left(\dfrac{r}{\beta_0}\right) - \dfrac{2M(r)}{\beta_0} =0.
\end{equation} 

Horizons are visualised in figure~\ref{fig:horizons} for a variety of $\beta_0^2\kappa_\infty$ values. These horizon plots span $(-\infty, +\infty)$ in $\beta_0\gamma_0$ and $[0, +\infty)$ in $r/\beta_0$, where each horizontal slice corresponds to a causal structure as specified by a given $\beta_0\gamma_0$ value. Yellow and red regions denote $\mathrm{T}$ and $\mathrm{S}$ regions respectively, with the solid black lines which segregate these two domains corresponding to $\widetilde{\Delta}^\mathrm{H}=0$. The grey horizontal band is $0\leq\beta_0\gamma_0\leq2/3$ where $-\infty< r_\mathrm{st}\leq0$, and hence this range of $\beta_0\gamma_0$ is deemed to yield unphysical photon sphere radii. 
The dashed or dotted blue lines denote locations of stable photon spheres~\eqref{eq:rst} in T or S regions. Note that the dotted blue lines traversing red S regions correspond to unphysical photon spheres, and hence solutions with this line going through S regions are also unphysical solutions. The solid blue line at $\beta_0\gamma_0=0$ may be taken to correspond to the GRS ($\beta_0^2\kappa_\infty=0$), GRSdS ($\beta_0^2\kappa_\infty>0$), or GRSAdS ($\beta_0^2\kappa_\infty<0$) spacetime. As expected from figure~\ref{fig:turner_map_inf}, there are no stable photon spheres present in $\mathrm{T}$ regions for $\beta_0\gamma_0<0$ if $\beta_0^2\kappa_\infty\geq0$ (figure~\ref{subfig:horz_k>0}), as seen from the lack of blue dashed lines in these spacetimes. 

This compactification of the causal structures in the MK metric not only allows us to clearly see the different regions of spacetime, but helps us identify different types of horizons as well. Taking the first panel of figure~\ref{subfig:horz_k=0} as a reference, we can identify naked timelike singularities for large $\beta_0\gamma_0$ values. For smaller positive values of $\beta_0\gamma_0$, we obtain a GR Reissner-Nordstrom-like structure, with an interior $\mathrm{H}_ \mathrm{C}$ (Cauchy horizon) and exterior $\mathrm{H}_\mathrm{E}$ (event horizon). These two domains are separated by a metric that resembles the extremal Reissner-Nordstrom metric, corresponding to the horizontal slice that is tangential to the maximum of the black curve. Contrarily, in the lower half of the same figure for negative $\beta_0\gamma_0$, we have the opposite phenomenon. For negative $\beta_0\gamma_0$ values close to zero, we have a GRSdS-like geometry, with an interior $\mathrm{H}_\mathrm{E}$ (event horizon) and an exterior $\mathrm{H}_\Lambda$ (cosmological horizon). As we then decrease the value of $\beta_0\gamma_0$, these two horizons move towards each other and annihilate, creating a single $\mathrm{H}_\mathrm{N}$, in what we may call a quasi-Nariai limit. Past this limit, there are no horizons, and we have naked spacelike singularities. 

Notably, when the shell is loaded, we see a change in the topology of horizons in all three cases. In figures~\ref{subfig:horz_k>0} and~\ref{subfig:horz_k<0}, this change occurs in the unphysical $r_\mathrm{st}<0$ region, and in these cases the region at $r\rightarrow+\infty$ remains invariant regardless of the strength of the load. This corresponds to the fact that the nonzero $\beta_0 \kappa_\infty$ dictates the cosmological curvature; as $\kappa_\infty$ is defined at $r\rightarrow+\infty$, it is natural that e.g. $\mathrm{T}$ regions at $r\rightarrow+\infty$ when  $f_0=0$ remain $\mathrm{T}$ regions at nonzero $f_0$ (figure~\ref{subfig:horz_k<0}), and likewise for $\mathrm{S}$ regions (figure~\ref{subfig:horz_k>0}). On the other hand, this same rule is not followed in figure~\ref{subfig:horz_k=0}; as $\beta_0^2\kappa_\infty=0$, the spacetime curvature at large $r$ is instead dictated by the $\gamma_\infty r$ term in the blackening factor. The linear coefficient jumps from $\gamma_0$ to $\gamma_\infty$ across the photon shell, and this jump is negative; as a result, for a large-enough load, a positive $\gamma_0$ may jump to a negative $\gamma_\infty$ after the shell. Therefore, a previously positive contribution from $\gamma(r) r$ may become negative, resulting in a drastic topological change in the horizon structure. This is clearly seen in the $f_0\neq0$ panels of figure~\ref{subfig:horz_k=0}. 

An increase in $f_0$ results in an overall expansion of $\mathrm{S}$ regions and decrease of $\mathrm{T}$ regions, which is also confirmed by figure~\ref{fig:B(r)_line}. These new $\mathrm{S}$ regions expand by surrounding the dashed blue lines corresponding to $r=r_\mathrm{st}$; this is clear from e.g. the $f_0=1$ plots from figure~\ref{subfig:horz_k=0} in the $\beta_0\gamma_0>0$ range and figure~\ref{subfig:horz_k<0} in the $\beta_0\gamma_0<0$ range. Therefore, we confirm that increasing the strength of load at the hypersurface generally decreases $B(r)$, and moves $B(r_\mathrm{st})$ closer to a horizon. 

Notably, there are significant changes in the locations of horizons and general horizon geometry. These changes are most visible in figure~\ref{subfig:horz_k=0}, where certain $\beta_0\gamma_0$ ranges which previously did not possess any horizon interior to $r=r_\mathrm{st}$, develop one $\mathrm{H}_\mathrm{X}$ where $r=r_\mathrm{st}$ collides with the $\widetilde{\Delta}^\mathrm{H}=0$ line. Immediately under this intersection for smaller values of $\beta_0\gamma_0$, a pair of horizons forms: an exterior $\mathrm{H}_\mathrm{E}$ and an interior $\mathrm{H}_\mathrm{C}$. In agreement with our definitions from section~\ref{subsec:horizons}, these two horizons are seen to enclose a nested $\mathrm{S}^-$ region, mimicking a spacetime structure not unlike a subextremal charged (Reissner-Nordstrom) or rotating (Kerr) black hole spacetime in GR. An example of such a nested $\mathrm{S}^-$ region is seen in the e.g. $f_0=10$ case of figure~\ref{subfig:horz_k<0}, where the photon sphere radius for $\beta_0\gamma_0>0$ is seen to be located in an $\mathrm{S}^-$ region that is bounded by a $\mathrm{H}_\mathrm{C}$ to its left at a smaller $r/\beta_0$ and an $\mathrm{H}_\mathrm{E}$ at a larger $r/\beta_0$. 

This highlights the key observations to be made from figure~\ref{fig:horizons}. There must be a critical loading threshold value of $f_0$ such that the photon sphere is precisely coincident with a horizon ($B(r_\mathrm{st})=0$). Therefore, for an $f_0$ value past this threshold, the shell finds itself in an $\mathrm{S}^-$ region bounded by the aforementioned pair of horizons such that $B(r_\mathrm{st})<0$, and the existence of a null matter shell at this radius becomes no longer physical. These two observations are further investigated in section~\ref{sec:threshold&above}.

\section{Maximal loading amplitude}\label{sec:threshold&above} 

Following section~\ref{sec:geometries_pert}, this section is subdivided into two parts. First, we investigate the maximum value permitted for $f_0$ (section~\ref{subsec:threshold}), and then elucidate the near-horizon geometry obtained at the decoupling limit when $f_0$ is at a maximum value (section~\ref{subsec:near-horizon}).

\subsection{Threshold amplitude}\label{subsec:threshold}

The above analysis, which is analogous to loading the photon sphere with null matter, suggests the existence of a maximum value of $f_0$. Indeed, requiring the photon-loaded ($f(r)\neq0$) blackening factor $B(r_\mathrm{st})>0$, we obtain: 
\begin{equation}\label{eq:threshold}
    f_0\leq \dfrac{6 B_0(r_\mathrm{st})}{r_\mathrm{st}^3}.
\end{equation}
Therefore, there is a threshold loading value of $f_0$ before $B(r_\mathrm{st})<0$, which is $f_\mathrm{max}=6 B_0(r_\mathrm{st})/{r_\mathrm{st}^3}$. As $\mathrm{d}B(r_\mathrm{st})/\mathrm{d}r=0$ for $f_0=f_\mathrm{max}$, the resulting horizon is classified as extremal ($\mathrm{H}_\mathrm{X}$). 

The nature of this $\mathrm{H}_\text{X}$ is fundamentally different to those found in usual extremal metrics, like the extremal charged or rotating metrics in GR~\cite{deFelice_2001}. $\mathrm{H}_\text{X}$ is usually created by the mutual annihilation of an exterior $\mathrm{H}_\text{E}$ with an interior $\mathrm{H}_\mathrm{C}$, when the charge or spin of the black hole reaches a critical value, such that surpassing this limit results in the black hole becoming a naked singularity unconcealed by any horizon, violating the weak cosmic censorship hypothesis~\cite{deFelice_2001}. Specifically, $\mathrm{H}_\text{E}$ and $\mathrm{H}_\mathrm{C}$ move towards each other and converge at the extremal limit. However, in our metric with a nonzero load at the photon sphere, $\mathrm{H}_\text{X}$ is not created through the annihilation of two counteracting horizons, but rather manifested as a result of the deformation of the blackening factor. In short, $\mathrm{H}_\mathrm{X}$ pops into existence independently of the surrounding horizon structure, as opposed to forming due to the coalescence of counteracting horizons~\cite{bekenstein1994}.

\subsection{\texorpdfstring{Near-horizon geometry at $\mathrm{H}_\mathrm{X}$}{Near-horizon geometry at H\_X}}\label{subsec:near-horizon}

We establish the near-horizon geometry at the new extremal horizon at the stable photon sphere, by following the procedures of~\cite{castro2023}. To do so, we may reformulate the blackening factor $B(r)$ in the following manner: 
\begin{equation}\label{eq:MK_NHG}
    B(r)=q(r)\left(1-\dfrac{r_\mathrm{st}}{r}\right)^2,
\end{equation}
where $q(r)$ is a quadratic. We consider this formulation to be valid, as $\mathrm{H}_\mathrm{X}$ is a double root of the metric placed exactly at $r=r_\mathrm{st}$. 

We first evaluate the near-horizon geometry of $\mathrm{H}_\mathrm{X}$ using the parameters of the MK metric, which we expect to be AdS$_2\times$S$^2$ based on previous works~\cite{castro2023} investigating metric geometries at similar limits in GR. 
To obtain the geometry at the decoupling limit, we consider the following coordinate transformation~\cite{bardeen1999}: 
\begin{align}
    r=r_\mathrm{st}+\lambda R, \hspace{1cm} t=\dfrac{T}{\lambda},
\end{align}
with $R$ and $T$ defining new radial and temporal variables respectively. $\lambda$ is the dimensionless decoupling parameter, which is set to zero in the decoupling limit to obtain a metric describing the geometry near $\mathrm{H}_\mathrm{X}$~\cite{castro2023}.

Using these new coordinates, we reformulate the line element and obtain an AdS$_2\times$S$^2$ geometry in the near-horizon limit, which is described by the following ``AdS-form'' metric~\cite{hartnoll2009}: 
\begin{align}\label{eq:metric_AdS2XS2} 
    \mathrm{d}s^2 &\approx -\dfrac{R^2}{\ell_\mathrm{A}^2}\mathrm{d}T^2 + \dfrac{\ell_\mathrm{A}^2}{R^2}\mathrm{d}R^2 + r_\mathrm{st}^2\mathrm{d}{\Omega_2}^2,
\end{align} 
with the AdS$_2$ radius $\ell_\mathrm{A}$ evaluated to be
\begin{align}
    \ell_\mathrm{A} = 3\beta_0 - 2/\gamma_0,
\end{align}
which is exactly equal to the S$^2$ radius $r_\mathrm{st}$. This contrasts the near-horizon geometry of extremal metrics with cosmological curvature in GR, such as the Reissner-Nordstrom (Anti-)de Sitter (GRRN(A)dS) metric, for which the blackening factor is: 
\begin{equation}\label{eq:RNAdS_extremal} 
    B(r)=1-\dfrac{2\beta}{r}+\dfrac{Q^2 G}{4\pi\epsilon_0r^2}- \dfrac{r^2}{\ell_4^2},
\end{equation} 
with $Q$ corresponding to the charge of the central black hole, and $\Lambda=3/\ell_4^2$ the cosmological constant. In the extremal case of~\eqref{eq:RNAdS_extremal}, the AdS$_2$ radius $\ell_\mathrm{A}$ at the extremal horizon radius $r_\mathrm{X}$ is~\cite{castro2023}: 
\begin{equation} 
    \ell_\mathrm{A}^2=\dfrac{r_\mathrm{X}^2}{1-\dfrac{6r_\mathrm{X}^2}{\ell_4^2}}. 
\end{equation} 

The contrast between these two extremal metrics is significant. For the GRRN(A)dS metric described in~\eqref{eq:RNAdS_extremal}, the cosmological curvature $\kappa$ has a clear contribution to the AdS$_2$ radius $\ell_\mathrm{A}$ in the near-horizon limit. However, in our metric, the AdS$_2$ radius has no such dependence on the cosmological curvature parameter, \textit{despite} the presence of a cosmological $\kappa r^2$ term in the blackening factor. 

Therefore, at the maximum loading amplitude $f_0=f_\mathrm{max}$, we create an extremal metric that may have cosmological curvature, but unlike in GR, this curvature has \textit{no effect} on the AdS$_2$ radius of the near-$\mathrm{H}_\mathrm{X}$ geometry. 

However, because our evaluation of the near-horizon geometry is not unique to the photon-loaded metric discussed in the present work, it is likely the case that the AdS$_2$ radius is equal to the S$^2$ radius for all nonrotating CG metrics at the extremal limit, hence implying that the near-horizon spacetime is entirely decoupled from the asymptotic curvature, whether it be for the MK metric studied in the present work or the CG Reissner-Nordstr\"{o}m metric~\cite{mannheim1991_1, kusano2025_CGRN}. This would be a remarkable result unseen in GR, and may arise due to the additional conformal symmetry of CG. In general, a further study of extremal CG black holes is warranted, and may have significant consequences for CG in the realm of QG~\cite{astefanesei2008, hristov2010}.

\section{Conclusions and further perspectives}\label{sec:conclusion} 

In the present work, we have investigated the accumulation of photons at the stable photon sphere of the conformal gravity Mannheim-Kazanas metric, through a toy problem that assumes that these photons build up an infinitely-thin shell. Our important findings are summarised as follows: 
\begin{itemize} 
    \item Section~\ref{subsec:pressure}: A thin shell source defined by a Dirac $\delta$ function $f(r)$ induces a jump in the radial pressure, unless its radius is equal to either of the photon sphere radii of the Mannheim-Kazanas metric. 
    \item Section~\ref{subsec:PS-pert} and section~\ref{subsec:horizons-pert}: Upon loading the stable photon sphere with photons by increasing the amplitude of $f(r)$, its area remains invariant and its radius stays constant. This highlights a previously unseen significance of photon sphere radii in CG's MK metric~\cite{mannheim2025_perscomm} and the increased stability of the Mannheim-Kazanas metric in comparison to some metrics in GR~\cite{cunha2025}. However, locations of horizons may change. 
    \item Section~\ref{subsec:threshold} and section~\ref{subsec:near-horizon}: At a critical loading amplitude for the thin shell, an extremal horizon is created at the stable photon sphere radius. Its near-horizon geometry is AdS$_2\times$S$^2$ as expected, but with an AdS$_2$ radius exactly equal to that of the $\mathrm{S}^2$ radius. In GR, this occurs only for asymptotically flat metrics, with a zero cosmological curvature; in our CG case, the quadratic cosmological term $-\kappa r^2$ has no effect on the AdS$_2$ radius, even if $\kappa$ is nonzero. 
\end{itemize} 

To conclude, we propose some potential directions in which the present work can be extended. 

First, we note that as our current formulation of the shell problem is an approximation, adding a finite width to the shell would be an interesting extension to the problem put forward in the present work. 

We also remark that while the independence of the near-horizon geometry to the cosmological curvature is an interesting result (section~\ref{subsec:near-horizon}), we suspect that this is not unique to the photon-loaded CG solution discussed in the present work. Indeed, if this result applies to sourceless Mannheim-Kazanas metrics as well, this would be a significant result. We aim to address this in the future, but deem this outside of the scope for the present work. 

Additionally, one may consider how the spacetime alters after the threshold loading limit~\eqref{eq:threshold} is exceeded. As this creates an interior Cauchy horizon and outer event horizon, the shell would fall inwards and subsequently backreact on the geometry; such a problem may be investigated in the future by adapting numerical relativity routines such as \textsc{GRChombo}~\cite{andrade2021} or \textsc{NRPy}~\cite{ruchlin2018} to solve the time-dependent CG field equations~\cite{kazanas1991}. 

Finally, we direct the readers' attention to a recent work by Foster et al.~\cite{foster2016}, which concerns a solution to the Einstein field equations~\eqref{eq:EFE} that describes a compact object consisting solely of orbiting photons. Formulating such a solution in CG would no doubt be a worthwhile and feasible endeavour, considering previous works which have tackled boson stars~\cite{brihaye2009}, cosmic strings~\cite{verbin2011}, and Dirac solitons~\cite{leggat2017} in conformal gravity. 

\nocite{*}
\section*{Acknowledgments} 
Parts of the present work adapt the MPhys dissertation project report of Reinosuke Kusano, cosupervised by coauthors Keith Horne and Friedrich Koenig. We thank Philip Mannheim for his time and inputs regarding the third-order constraint problem. 

\section*{References}

\bibliography{Bibliography}

\end{document}